\documentclass[10pt,a4paper]{article}

\usepackage[margin=2cm]{geometry}
\usepackage{amsfonts}
\usepackage{graphicx} 
\usepackage{subfig}
\usepackage{cite}
\usepackage{caption}
\usepackage{amsmath}
\usepackage{amssymb}
\usepackage{color}
\usepackage{soul}
\usepackage[normalem]{ulem}
\usepackage{float}
\usepackage{authblk}  
\usepackage[utf8]{inputenc}
\usepackage{epsfig}
\usepackage{float} 
\usepackage{tikz}
\usetikzlibrary{patterns}
\usepackage[toc,page]{appendix}

\definecolor{DarkGreen}{RGB}{0,215,0}

\title{\textbf{Epicyclic Frequencies for a Generic Ultracompact Object}}
\author[1,2]{Jorge F. M. Delgado\footnote{jorgedelgado@ua.pt}}
\affil[1]{\normalsize Departamento de Matemática da Universidade de Aveiro and 

Center for Research and Development in Mathematics and Applications -- CIDMA

Campus de Santiago, 3810-183 Aveiro, Portugal}
\affil[2]{\normalsize Centro de Astrofísica e Gravitação - CENTRA,

Departamento de Física, Instituto Superior Técnico - IST,

Universidade de Lisboa - UL,

Av. Rovisco Pais 1, 1049-001 Lisboa, Portugal}

\begin{document}

\maketitle

\begin{abstract}
\normalsize

Recently, it has been shown that the radial stability of a light-ring (LR) in a spacetime generated by a stationary, axisymmetric, asymptotically flat object with a $\mathbb{Z}_2$ symmetry determines the possibility and radial stability of timelike circular orbits (TCOs) around the LR. 
In this paper, we generalise this result by also considering the vertical (angular) stability of the orbits through the study of the radial and vertical epicyclic frequencies. 
We show that the vertical stability of the LR only determines the vertical stability of the TCOs around it.  
A relation between the sum of the squared epicyclic frequencies and the Ricci tensor is also provided. 
With such relation, we show that objects with radially and vertically unstable LRs (TCOs) violate the null (strong) energy condition. 

\end{abstract}

\tableofcontents

\section{Introduction}

The study of the structure of null and timelike circular orbits around a given generic ultracompact object is important to access several of its phenomenology properties. These properties can, in principal, be very different from the ones found for the paradigmatic General Relativity black hole (BH) -- the Kerr BH \cite{Kerr:1963ud}. 
Such differences are opportune to know given the recent measurements of gravitational waves (GWs) by the LIGO/Virgo/KAGRA collaboration~\cite{LIGOScientific:2016aoc,LIGOScientific:2018mvr,LIGOScientific:2020ibl,LIGOScientific:2021djp} and the shadow and lensing of light around the M87* and Sgr A* supermassive BHs provided by the EHT collaboration~\cite{EventHorizonTelescope:2019dse,EventHorizonTelescope:2022xnr}. 
Furthermore, with the advances towards the construction and implementation of the future Laser Interferometer Space Antenna (LISA), designed to be able to detect GW signals with much lower frequencies than the LIGO/Virgo/KAGRA collaboration, such as the ones produced by extreme mass ratio inspirals (EMRIs)~\cite{Barausse:2020rsu}, it will be possible to probe the structure of timelike circular orbits around these ultracompact objects, that can be BHs or other ultracompact objects that could mimic a BH -- \textit{eg.} scalar and vector boson stars \cite{PhysRev.172.1331,PhysRev.187.1767,Schunck_2003,Herdeiro:2017fhv,Brito:2015pxa,Herdeiro:2019mbz,Bustillo:2020syj,Herdeiro:2021lwl}.

To get a sense of what structures of circular orbits one can find, we start by recalling three remarkable works developed recently in~\cite{Cunha:2017qtt,Cunha:2020azh,Ghosh:2021txu}. 
In the first work~\cite{Cunha:2017qtt}, the authors showed, using a topological argument, that a generic stationary, axisymmetric and asymptotically flat ultracompact horizonless object must have at least two null circular orbits -- otherwise known as light-rings (LRs) --, where one of them is stable. 
In the second work~\cite{Cunha:2020azh}, the authors used a similar topological argument as in the previous one to prove that a stationary, axisymmetric, asymptotically flat and non-extremal BH must have at least one unstable LR outside of its event horizon. 
In the last work~\cite{Ghosh:2021txu}, it was shown that by using the previous two results together with a careful analyse of the boundary behaviour of some key quantities, it is possible to further conclude that, for both types of ultracompact objects mentioned (with or without an horizon), if the object possesses an ergoregion then, at least, one LR must exist outside its ergoregion.
All these results define precisely the structure of null circular orbits for generic (within the assumptions mentioned) ultracompact objects, with or without an horizon. 

A natural question to ask now is the following: does the existence of a LR establish any structure for timelike circular orbits (TCOs) around it?
An answer to this question was given in \cite{Delgado:2021jxd}, where it was shown that, for a generic ultracompact object with the same assumptions as mentioned previously, together with a $\mathbb{Z}_2$ symmetry fixing an equatorial plane, the radial stability of the existing LRs determines the localisation and radial stability of TCOs in their vicinity. However, this work only mentions and studies the radial stability of the circular orbits, and it leaves open an opportunity to perform a more complete study where the vertical stability of the circular orbits is also included.

In this paper, we shall take the aforementioned opportunity.
For that, we will study the radial and vertical epicyclic frequencies of null and timelike particles following circular orbits on a spacetime generated by a generic ultracompact object with the same assumptions mentioned in the previous paragraph.
These epicyclic frequencies are computed by perturbing the circular orbits in the radial and vertical directions, and are tightly connected with the stability of the orbits.
With them, we shall arrive at the same result present in \cite{Delgado:2021jxd}, together with its generalisation, where we also look at the vertical stability of the orbits.




Throughout the first part of this paper, we do not impose any further assumption upon the ultracompact object in question. The generic ultracompact object may or may not be a BH, and it is a solution of an undefined set of equations of motion obtained from an undefined theory of gravity. However, it is expected that the matter which composes the object obeys the energy conditions. Therefore, in the second part of this paper, we establish connections between the epicyclic frequencies, and consequently the structure of circular orbits, with the energy conditions.

The first study of its kind was done in~\cite{10.1093/mnras/stt1185}, where the authors found that the sum of the squared epicyclic frequencies of TCOs around a Maclaurin spheroid~\cite{maclaurin1742treatise,chandrasekhar1969ellipsoidal} in the Newtonian regime, could be written only in terms of the angular velocity of the timelike particles, $\Omega$, and the density of the spheroid, $\rho$, as~\cite{10.1093/mnras/stt1185},
\begin{equation}\label{Eq:SumEpicyclicFreqNewtonian}
	\omega_r^2 + \omega_\theta^2 = 2 \Omega^2 + 4\pi G \rho~,
\end{equation}
where $\omega_r$ and $\omega_\theta$ are the radial and vertical epicyclic frequencies, respectively, and $G$ is Newton's constant. 
In the same work, the authors also found that, for the case of a Kerr BH, the sum vanishes when computed at the LR.

Some years later, a proof developed in~\cite{Vieira:2017zau} showed that, for a very generic static, axisymmetric and asymptotically flat object, the sum of the squared epicyclic frequencies of TCOs could be written at the expense of a linear combination of the components of the Ricci tensor~\cite{Vieira:2017zau},
\begin{equation}\label{Eq:VieiraResult}
	\omega_r^2 + \omega_\theta^2 = \left[ R_{tt} + \Omega^2 R_{\varphi\varphi} \right] + \frac{\Omega^2}{2g_{rr}}  \frac{d g_{\varphi\varphi}}{d r} \frac{1}{\tilde{r}^2} \frac{d\tilde{r}^2}{dr^2}~,
\end{equation}
where, $\tilde{r}^2 = -g_{\varphi\varphi}/g_{tt}$. They also found, through the above equation, that if the object obeys the strong energy condition (SEC) then $\omega_r^2 + \omega_\theta^2 > 0$.

In the second part of this paper, we shall generalise the work done in~\cite{Vieira:2017zau} by considering a more generic stationary ultracompact object. We shall obtain a similar result to the one presented in Eq. \eqref{Eq:VieiraResult} and we shall consider the SEC as well as the null energy condition (NEC). If the matter which compose the object in question obeys the SEC and NEC, then the sum of the epicyclic frequencies of LRs and TCOs will always be non negative.


This work is organised as follows. 
In Section 2, we start by introduction and defining all quantities of interest for the null and timelike circular orbits, namely, their radial and vertical epicyclic frequencies. 
Then, in Section 3, we show, together with the work done in \cite{Delgado:2021jxd}, that the epicyclic frequencies of the timelike particle coincide with the epicyclic frequencies of the null particles when both of them are computed at the LRs. The four possible structure of TCOs surrounding a LR are presented.
Section 4 provides the relation between the sum of the squared epicyclic frequencies and the Ricci tensor. Connections with the null and strong energy condition are also provided and analysed. 
Finally, in Section 5, we close the present paper with the conclusions and final remarks.

\section{Epicyclic frequencies of equatorial circular orbits}
\label{Sec:Definitions}

To study the epicyclic frequencies, we shall follow a similar setup as the one described in \cite{Delgado:2021jxd}. Let us consider a stationary, axisymmetric, asymptotically flat, 1+3 dimensional spacetime, $(\mathcal{M}, g)$, that describes a generic ultracompact object with a $\mathbb{Z}_2$ symmetry that may or may not have an event horizon. No assumption is made on the field equations $(\mathcal{M},g)$ solves.

Stationarity and axial symmetry imply the existence of two Killing vectors field, $\{\eta_1, \eta_2\}$, that commune, $[\eta_1, \eta_2] = 0$, thanks to a theorem develop by Carter (together with asymptotic flatness) \cite{Carter:1970ea}. 
Such result open the possibility to choose a coordinate system, $(t,r,\theta,\varphi)$, that can be adapted to the Killing vector fields such that $\eta_1 = \partial_t$ and $\eta_2 = \partial_\varphi$. 
We also assume that the metric associated to the spacetime is, at least, $C^2$-smooth on and outside the possible horizon, and circular. This implies, for asymptotically flat spacetimes, that the geometry possesses a 2-space orthogonal to the Killing vector fields (\textit{c.f.} theorem 7.11 in \cite{Wald:1984rg}). Therefore, the discrete symmetry $(t,\varphi) \rightarrow (-t,-\varphi)$ is present on the spacetime.

By a gauge choice, we can defined spherical-like coordinates $(r,\theta)$ in the orthogonal 2-space that are orthogonal to each other. An extra gauge choice can be used to fix the localisation of the horizon at a constant positive radial coordinate, $r = r_H$, for the case of ultracompact objects with a horizon. With these choices, $g_{r\theta} = 0$, $g_{rr} > 0$ and $g_{\theta\theta} > 0$ (outside the possible horizon). One can also impose that $(r,\theta)$ must reduce to the standard spherical coordinates as one approaches spatial infinity, $r \rightarrow \infty$. The range of the coordinates are, $t \in (-\infty, +\infty)$; $r \in (r_H,+\infty)$, if a horizon exists, or $r \in [0, +\infty)$, if no horizon exists; $\theta \in [0, \pi]$; and $\varphi \in [0, 2\pi)$. The rotating axis is located at $\theta = \{0,\pi\}$, and the equatorial plane can be found at $\theta = \pi/2$. Outside of a possible horizon, causality implies that $g_{\varphi\varphi} \geq 0$.

In the end, due to all the assumptions, gauge choices, and using a Lorentzian signature $(-,+,+,+)$, we can write the following line element,
\begin{equation}
	ds^2 = g_{tt}(r,\theta) dt^2 + 2 g_{t\varphi}(r,\theta) dt d\varphi + g_{\varphi\varphi}(r,\theta) d\varphi^2 + g_{rr}(r,\theta) dr^2 + g_{\theta\theta}(r,\theta) d\theta^2 ~.
\end{equation}
Note that we shall consider that the radial coordinate is a faithful measurement of the distance to the ultracompact object.

The motion of test particles in the above geometry can be described through the following effective Lagrangian,
\begin{equation}
	2 \mathcal{L} = g_{\mu\nu} \dot{x}^\mu \dot{x}^\nu = \epsilon ~,
\end{equation}
where the dot denotes the derivation with respect to an affine parameter, and $\epsilon = \{-1, 0\}$ for timelike and null particles, respectively. 
The effective Lagrangian will depend on the spherical-like coordinate, and can be written as follows,
\begin{equation}
	2 \mathcal{L} = 	g_{tt}(r,\theta) \dot{t}^2 + 2 g_{t\varphi}(r,\theta) \dot{t} \dot{\varphi} + g_{\varphi\varphi}(r,\theta) \dot{\varphi}^2 + g_{rr}(r,\theta) \dot{r}^2 + g_{\theta\theta}(r,\theta) \dot{\theta}^2 = \epsilon ~.
\end{equation}

The existence of Killing vector fields give raise to constants of motion that we can be introduce into the Lagrangian. Those are the energy, $E$, and angular momentum, $L$, of the test particle,
\begin{equation}
	-E \equiv g_{t\mu} \dot{x}^\mu = g_{tt} \dot{t} + g_{t\varphi} \dot{\varphi} ~, \hspace{10pt} L \equiv g_{\varphi \mu} \dot{x}^\mu = g_{\varphi t} \dot{t} + g_{\varphi\varphi} \dot{\varphi} ~.
\end{equation}
The effective Lagrangian reads now,
\begin{equation}\label{Eq:LagrangianAB}
	2 \mathcal{L} = - \frac{A(r,\theta,E,L)}{B(r,\theta)} + g_{rr}(r,\theta) \dot{r}^2 + g_{\theta\theta}(r,\theta) \dot{\theta}^2= \epsilon ~, 
\end{equation}
where
\begin{equation}
	A(r,\theta,E,L) = g_{\varphi\varphi}(r,\theta) E^2 + 2 g_{t\varphi}(r,\theta) E L + g_{tt}(r,\theta) L^2 ~, \hspace{10pt} \text{and} \hspace{10pt} B(r,\theta) = g_{t\varphi}(r,\theta)^2 - g_{tt}(r,\theta) g_{\varphi\varphi}(r,\theta) ~.
\end{equation}
Eq. \eqref{Eq:LagrangianAB} suggests the introduction of an effective potential, $V_\epsilon(r,\theta)$, defined in the following way,
\begin{equation}
	V_\epsilon(r,\theta) \equiv g_{rr}(r,\theta) \dot{r}^2 + g_{\theta\theta}(r,\theta) \dot{\theta}^2= \epsilon + \frac{A(r,\theta,E,L)}{B(r,\theta)} ~.
\end{equation}

In this work, we are interested in objects with a $\mathbb{Z}_2$ symmetry and in circular orbits in the equatorial plane, $\theta = \pi/2$, meaning that $\dot{\theta} = 0$. With these additional assumptions, the effective potential will only have a radial dependency, and can be written as,
\begin{equation}
	V^r_\epsilon(r) \equiv g_{rr}(r,\pi/2) \dot{r}^2 = \epsilon + \frac{A(r,\pi/2,E,L)}{B(r,\pi/2)} ~.
\end{equation}

An equatorial circular orbit can be easily computed through this new effective potential, by imposing that the following two equations must be true simultaneously, 
\begin{equation}\label{Eq:FirstEquationPotential}
	V^r_\epsilon(r_\text{cir}) = 0 \hspace{20pt} \Leftrightarrow \hspace{20pt} A(r_\text{cir},\pi/2,E,L) = - \epsilon B(r_\text{cir},\pi/2) ~,
\end{equation}
and
\begin{equation}\label{Eq:SecondEquationPotential}
	\partial_r V^r_\epsilon(r_\text{cir}) = 0 \hspace{20pt} \Leftrightarrow \hspace{20pt} \partial_r A(r_\text{cir},\pi/2,E,L) = -\epsilon \partial_r B(r_\text{cir},\pi/2) ~.
\end{equation}
Note that we have used Eq. \eqref{Eq:FirstEquationPotential}, to obtain the Eq. \eqref{Eq:SecondEquationPotential}.

Beside knowing where and how to find circular orbits, it is equally important to know their stability. This can be analysed by computing both their radial and vertical epicyclic frequencies.

\subsection{Radial and vertical epicyclic frequencies}

The epicyclic frequencies can be computed by perturbing a circular orbit in either the radial or vertical direction. Through the former, we compute the radial epicyclic frequency, whereas through the latter, we obtain the vertical epicyclic frequency.

Let us start by consider that $x$ is one of the spheroidal coordinates, $x = \{r,\theta\}$, and $y$ is the remaining spheroidal coordinate. Moreover, let us also consider that we are on a circular orbit, such that $x = x_\text{c}$ and $y = y_\text{c}$. We shall fix $y$ and introduce a perturbation in $x$ around $x_\text{c}$, such that $y = y_\text{c} \Leftrightarrow \dot{y} = 0$ and $x = x_\text{c} + \delta x  \Leftrightarrow \dot{x} = \dot{\delta x}$. Under these assumptions, the perturbed effective potential reads,
\begin{equation}\label{Eq:PerturbedPotential}
	V^x_\epsilon(x_\text{c} + \delta x,y_\text{c}) = g_{xx}(x_\text{c} + \delta x,y_\text{c}) \dot{\delta x}^2 ~.
\end{equation}
The left-hand side of the above equation can be expanded to,
\begin{equation}
	V^x_\epsilon(x_\text{c} + \delta x,y_\text{c}) = V^x_\epsilon(x_\text{c},y_\text{c}) + \partial_x V^x_\epsilon(x_\text{c},y_\text{c}) \delta x + \frac{1}{2} \partial_x^2 V^x_\epsilon(x_\text{c},y_\text{c}) \delta x^2 + \mathcal{O}(\delta x^3) ~.
\end{equation}
The first two terms will always vanish because we are both on a circular orbit (if $x=r: V^r_\epsilon(r_\text{cir}) = \partial_r V^r_\epsilon(r_\text{cir}) = 0$) and on the equatorial plane of a $\mathbb{Z}_2$ symmetric object (if $x=\theta: V^\theta_\epsilon(\pi/2) = \partial_\theta V^\theta_\epsilon(\pi/2) = 0$). Thus, equation \eqref{Eq:PerturbedPotential} becomes\footnote{Note that, for simplicity and notation ease, we dropped the error term, $\mathcal{O}(\delta x^3)$.},
\begin{equation}
	\frac{1}{2} \partial^2_x V^x_\epsilon(x_\text{c},y_\text{c}) \delta x^2 = g_{xx}(x_\text{c} + \delta x,y_\text{c}) \dot{\delta x}^2 ~.
\end{equation}
Expanding now the right-hand size of the above equation, and considering only terms up to second order in $\dot{\delta x}$, we arrive at,
\begin{equation}
	\frac{1}{2} \partial^2_x V^x_\epsilon(x_\text{c},y_\text{c}) \delta x^2 = g_{xx}(x_\text{c},y_\text{c}) \dot{\delta x}^2 ~.
\end{equation}
By performing a dot derivative of both sizes, we can write a more instructive result,  
\begin{equation}
	\ddot{\delta x} + (\omega^x_\epsilon)^2 \delta x = 0  ~,\hspace{20pt} \text{where} \hspace{20pt} (\omega^x_\epsilon)^2 \equiv -\frac{1}{2} \frac{\partial_x^2 V^x_\epsilon(x_\text{c},y_\text{c})}{g_{xx}(x_\text{c},y_\text{c})}
\end{equation}
This is an harmonic oscillator, which means that if the frequency $\omega_x^2$ is negative (positive), the perturbation grows exponentially (remains a small perturbation) leading to unstable (stable) circular orbits. This frequency is known as the epicyclic frequency. 
Going back to the spheroidal coordinates, we have the radial and vertical epicyclic frequencies, respectively,
\begin{equation}\label{Eq:EpicyclicFrequenciesComovingObs}
	(\omega^r_\epsilon)^2 \equiv -\frac{1}{2} \frac{\partial_r^2 V^r_\epsilon(r_\text{cir},\pi/2)}{g_{rr}(r_\text{cir},\pi/2)} ~, \hspace{20pt} (\omega^\theta_\epsilon)^2 \equiv -\frac{1}{2} \frac{\partial_\theta^2 V^\theta_\epsilon(r_\text{cir},\pi/2)}{g_{\theta\theta}(r_\text{cir},\pi/2)} ~.
\end{equation}
These epicyclic frequencies are measured with respect to the proper time of a comoving observer. However, in the following computations, it proved useful to use the epicyclic frequencies measured by an observer at spatial infinity, rather than by a comoving observer.
To obtain the frequencies measured by such observer, one only has to divide the frequencies in Eq. \eqref{Eq:EpicyclicFrequenciesComovingObs} by the squared redshift factor, $\dot{t}^2$,
\begin{equation}\label{Eq:EpicyclicFrequencyRadial}
	(\nu^r_\epsilon)^2 \equiv \frac{(\omega^r_\epsilon)^2}{\dot{t}^2} = -\frac{1}{2} \frac{\partial_r^2 V^r_\epsilon(r_\text{cir},\pi/2)}{g_{rr}(r_\text{cir},\pi/2)} \left[ \frac{B(r_\text{cir},\pi/2)}{E g_{\varphi \varphi}(r_\text{cir},\pi/2) + L g_{t\varphi}(r_\text{cir},\pi/2) }\right]^2 ~,~
\end{equation}
\begin{equation}\label{Eq:EpicyclicFrequencyAngular}
	(\nu^\theta_\epsilon)^2 \equiv \frac{(\omega^\theta_\epsilon)^2}{\dot{t}^2} = -\frac{1}{2} \frac{\partial_\theta^2 V^\theta_\epsilon(r_\text{cir},\pi/2)}{g_{\theta\theta}(r_\text{cir},\pi/2)} \left[ \frac{B(r_\text{cir},\pi/2)}{E g_{\varphi \varphi}(r_\text{cir},\pi/2) + L g_{t\varphi}(r_\text{cir},\pi/2) }\right]^2 ~.
\end{equation}
Henceforth, we shall drop the explicit dependency of the several functions, and it shall be understood that all quantities are computed at $r=r_\text{cir}$ and $\theta = \pi/2$, unless stated otherwise.

We shall now compute the epicyclic frequencies for null and timelike circular orbits.

\subsection{Null particles}

For null particles, $\epsilon = 0$, circular orbits are known as LRs. 
In order to obtain them, we will use Eqs. \eqref{Eq:FirstEquationPotential} and \eqref{Eq:SecondEquationPotential}. 
Both equations can be rewritten in terms of the inverse impact parameter, $\sigma_\pm = E_\pm/L_\pm$, where $\pm$ represents the two possible solutions due to the rotation of the ultracompact object and they are associated with prograde ($+$) and retrograde ($-$) orbits,
\begin{eqnarray}
	A = 0 \hspace{10pt} &\Leftrightarrow& \hspace{10pt} \left[ g_{\varphi\varphi} \sigma_\pm^2 + 2 g_{t\varphi} \sigma_\pm + g_{tt} \right]_\text{LR} = 0~,\label{Eq:LRFirstEquationPotential} \\
	\partial_r A = 0 \hspace{10pt} &\Leftrightarrow& \hspace{10pt} \left[\partial_r g_{\varphi\varphi} \sigma_\pm^2 + 2 \partial_r g_{t\varphi} \sigma_\pm + \partial_r g_{tt} \right]_\text{LR} = 0 ~.\label{Eq:LRSecondEquationPotential}
\end{eqnarray}
The first equation gives an algebraical equation for the inverse impact parameter,
\begin{equation}
	\sigma_\pm = \left[ \frac{-g_{t\varphi} + \sqrt{B}}{g_{\varphi\varphi}} \right]_\text{LR}~,
\end{equation}  
whereas the second equation gives the radial coordinate of the LR, $r = r_\text{LR}$.

The epicyclic frequencies of the LRs, can be easily computed through Eqs. \eqref{Eq:EpicyclicFrequencyRadial} and \eqref{Eq:EpicyclicFrequencyAngular},
\begin{equation}
	(\nu^r_{0})^2 = - \frac{1}{2} \left[ \frac{\partial_r^2 g_{\varphi\varphi} \sigma_\pm^2 + 2 \partial_r^2 g_{t\varphi} \sigma_\pm + \partial_r^2 g_{tt}}{g_{rr}} \right]_\text{LR}
\end{equation}
\begin{equation}
	(\nu^\theta_{0})^2 = - \frac{1}{2} \left[ \frac{\partial_\theta^2 g_{\varphi\varphi} \sigma_\pm^2 + 2 \partial_\theta^2 g_{t\varphi} \sigma_\pm + \partial_\theta^2 g_{tt}}{g_{\theta\theta}} \right]_\text{LR}
\end{equation}
Since $g_{rr}$ and $g_{\theta\theta}$ are always positive outside of a possible horizon, the epicyclic frequencies are real (complex) if their numerator are negative (positive).

\subsection{Timelike particles}

For timelike particles, $\epsilon = -1$, the same analysis can be done. In order to simplify the following computations, it is convenient to introduce the angular velocity of timelike particles along circular orbits (measured with respect to an observer at infinity),
\begin{equation}
	\Omega = \frac{d\varphi}{dt} = - \frac{E g_{t\varphi} + L g_{tt}}{E g_{\varphi\varphi} + L g_{t\varphi}} ~.
\end{equation}
With this result and Eq. \eqref{Eq:FirstEquationPotential}, one can obtain an analytical expression for the energy and angular momentum of the timelike particle written in terms of the metric functions and the angular velocity,
\begin{equation}\label{Eq:EnergyAngMomTCOs}
	E_\pm = - \left[ \frac{g_{tt} + g_{t\varphi}\Omega_\pm}{\sqrt{\beta_\pm}} \right]_{r_\text{cir}}~,\hspace{10pt} L = \left[ \frac{g_{t\varphi} + g_{\varphi\varphi}\Omega_\pm}{\sqrt{\beta_\pm}} \right]_{r_\text{cir}} ~,
\end{equation}
where $\beta_\pm \equiv -g_{tt} - 2 g_{t\varphi} \Omega_\pm - g_{\varphi\varphi} \Omega_\pm^2$. 
An expression for the angular velocity can be obtained by solving equation \eqref{Eq:SecondEquationPotential},
\begin{equation}\label{Eq:AngularVelTCOs}
	\Omega_\pm = \left[ \frac{-\partial_r g_{t\varphi} \pm \sqrt{C}}{\partial_r g_{\varphi\varphi}} \right]_{r_\text{cir}} ~,
\end{equation}
where $C \equiv (\partial_r g_{t\varphi})^2 - \partial_r g_{tt} \partial_r g_{\varphi\varphi}$.

With Eqs. \eqref{Eq:EnergyAngMomTCOs} and \eqref{Eq:AngularVelTCOs}, we have defined all possible timelike circular orbits (TCOs) in a given spacetime (with the assumptions presented in the beginning of this section). Their stability can be analysed through their epicyclic frequencies, 
\begin{equation}
	(\nu^r_{-1})^2 = - \frac{1}{2} \left[ \frac{(\partial_r^2 g_{\varphi\varphi} \Omega_\pm^2 + 2 \partial_r^2 g_{t\varphi} \Omega_\pm + \partial_r^2 g_{tt})B - 2 C \beta_\pm}{B g_{rr}} \right]_{r_\text{cir}} ~,
\end{equation}
\begin{equation}
	(\nu^\theta_{-1})^2 = - \frac{1}{2} \left[ \frac{\partial_\theta^2 g_{\varphi\varphi} \Omega_\pm^2 + 2 \partial_\theta^2 g_{t\varphi} \Omega_\pm + \partial_\theta^2 g_{tt}}{g_{\theta\theta}} \right]_{r_\text{cir}} ~.
\end{equation}
Similar as to the case of LRs, the epicyclic frequencies of TCOs are real (complex) is the numerator of the above expressions are negative (positive), since $B$, $g_{rr}$ and $g_{\theta\theta}$ are always positive outside a possible horizon.

\section{Connection between null and timelike particles}

We shall show now that the epicyclic frequencies of TCOs coincide with the epicyclic frequencies of LRs when we compute the former on a LR. 

\begin{figure}[h!]
	\centering
	\includegraphics[scale=0.21]{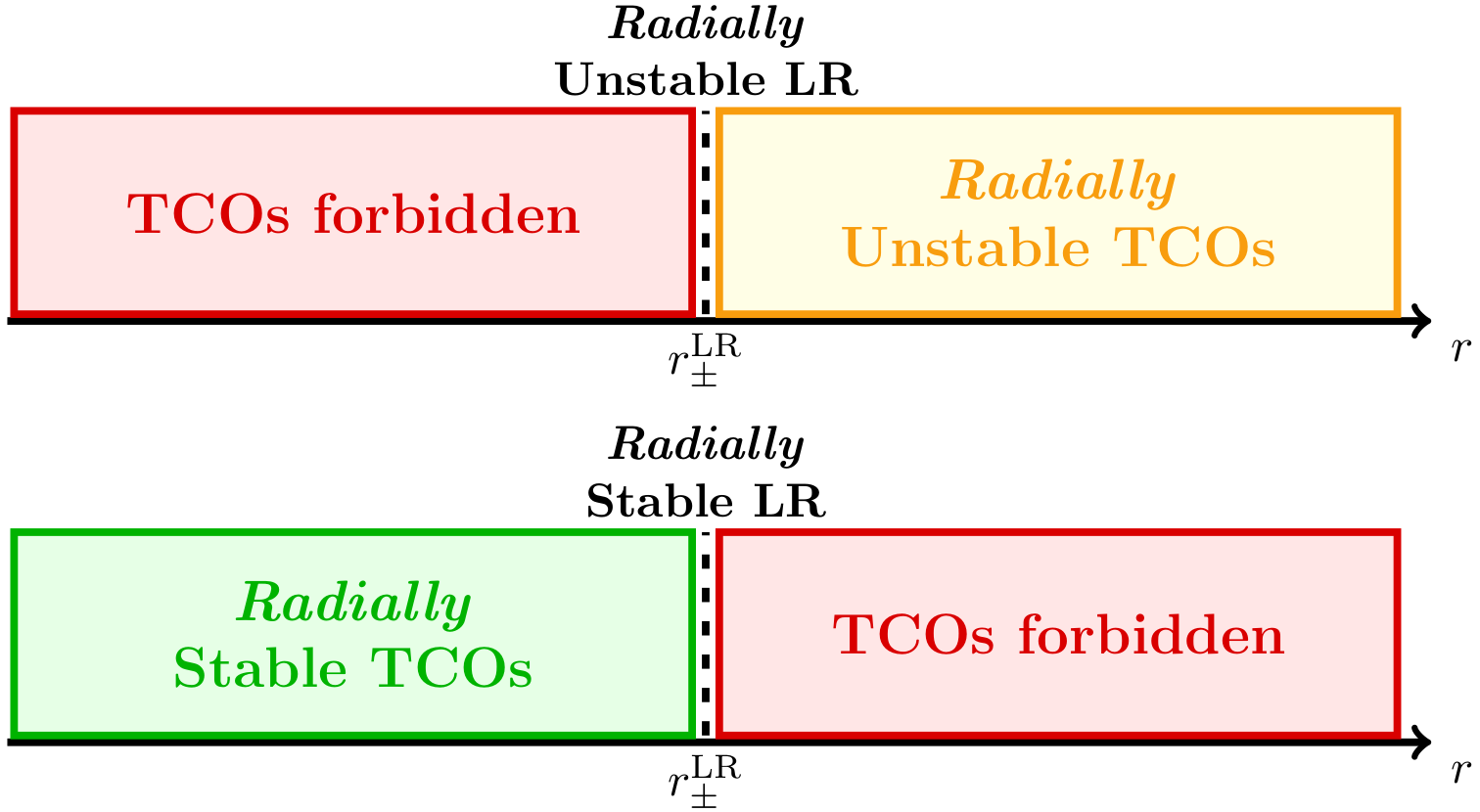}
	\caption{Structure of the equatorial TCOs in the immediate vicinity of an unstable (top panel) and stable (bottom panel) LR. Adapted from \cite{Delgado:2021jxd}.}
	\label{Fig:CompIllustStableLR}
\end{figure}

We start by recalling the results derived in \cite{Delgado:2021jxd}. Firstly, the authors showed that the function $\beta_\pm$ always vanishes on a LR, by proving that Eqs. \eqref{Eq:LRFirstEquationPotential} and \eqref{Eq:LRSecondEquationPotential} identically vanish when $\beta_\pm = 0$. A consequence of this result is that the angular velocity of a timelike particle on such a TCO is the same as the inverse impact parameter of a null particle on the LR, $[\Omega_\pm = \sigma_\pm]_\text{LR}$.
Secondly, they showed that the localisation and \textit{radial} stability of the regions that can harbour TCOs around the LR, depend exclusively on the \textit{radial} stability of said LR. Such statement is verified by looking at the sign of $\beta_\pm$ to find the regions where TCOs are allowed, and by studying the stability of those allowed regions. In short, a radially stable (unstable) LR accommodates radially stable (unstable) TCOs in the regions radially below (above) the LR. 
A schematic representation of these results is presented in Fig. \ref{Fig:CompIllustStableLR}.


These results can now pave the way for us to generalise them, by also including the vertical stability of the allowed TCOs.  For that, let us consider a spacetime with a LR. If we compute the epicyclic frequencies of a TCO infinitely close to the LR, such that we can use $\beta_\pm|_\text{LR} = 0$ and $[\Omega_\pm = \sigma_\pm]_\text{LR}$, we obtain,  

\begin{equation}\label{Eq:EpicyclicFreqRadialLR}
	(\nu^r_{-1})^2 = - \frac{1}{2} \left[ \frac{\partial_r^2 g_{\varphi\varphi} \sigma_\pm^2 + 2 \partial_r^2 g_{t\varphi} \sigma_\pm + \partial_r^2 g_{tt}}{g_{rr}} \right]_\text{LR} = (\nu^r_{0})^2 ~,
\end{equation}
\begin{equation}\label{Eq:EpicyclicFreqAngularLR}
	(\nu^\theta_{-1})^2 = - \frac{1}{2} \left[ \frac{\partial_\theta^2 g_{\varphi\varphi} \sigma_\pm^2 + 2 \partial_\theta^2 g_{t\varphi} \sigma_\pm + \partial_\theta^2 g_{tt}}{g_{\theta\theta}} \right]_\text{LR} = (\nu^\theta_{0})^2 ~.
\end{equation}
Here we see that the epicyclic frequencies, and hence the stability of a TCO infinitely close to a LR are precisely the same as the ones for the same LR. 
Therefore, the full stability (both radial and vertical) of a LR determines the full stability of a TCO infinitely close to that LR.
 
This result, however, it is only true infinitely close to the LR, where $\beta_\pm|_\text{LR} = 0$ and $[\Omega_\pm = \sigma_\pm]_\text{LR}$ are valid. To analyse the stability of the allowed TCOs in the vicinity of the LR, where $\beta_\pm|_\text{LR} = 0$ and $[\Omega_\pm = \sigma_\pm]_\text{LR}$ are no longer valid, we need to investigate what happens to the epicyclic frequencies on those regions. 
Fortunately, this analyse is rather simple to do. One only has to use the continuity properties of the epicyclic frequencies. 

Imagine a spacetime with a radial unstable LR. By Eq. \eqref{Eq:EpicyclicFreqRadialLR} we know that a timelike particle on a TCO infinitely close to the LR is radially unstable, hence, $\nu^r_{-1}(r_\text{LR})^2 < 0$. If we analyse the epicyclic frequency of an allowed TCO adjacent to the LR, one can argue that the epicyclic frequency squared will continue to be negative (however one does not know if it increases or decreases). Such argument can be done by assuming that the epicyclic frequencies are continuous on the allowed regions for TCOs, which is a fair assumption to make since we assumed that all metric functions are, at least, $C^2$-smooth and all remaining quantities are well-behaved. Hence, the allowed region of TCOs adjacent to the radially unstable LR will harbour radially unstable TCOs. Such result is consistent with the results provided in \cite{Delgado:2021jxd}. 

The same argument can be done regarding a radially stable LR, recovering the same results found in \cite{Delgado:2021jxd}, or regarding a vertically stable and unstable LR. In the last two cases, the allowed region of TCOs continues to the determined by the \textit{radial} stability of the LR, but their vertical stability is entirely determined by the vertical stability of the LR. We can summarise all the main results in the following list:
\begin{itemize}
	\item Spacetime with a \textit{radially} unstable LR.
	
	For such spacetime, the region immediately inwards of the LR can not accommodate TCOs, whereas the outward region can. To know the full stability of the allowed TCOs we need to know the full stability of the LR, therefore, let us assume that the LR is,
	\begin{itemize}
		\item \textit{Vertically} unstable.
			
			In this case, the allowed TCOs will be both \textit{radially} and \textit{vertically} unstable.
		
			\begin{figure}[H]
				\centering
				\includegraphics[scale=0.205]{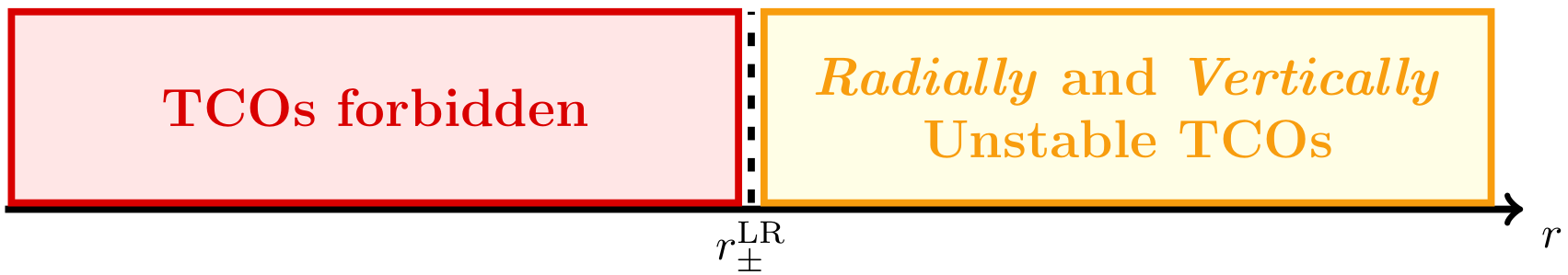}
				\caption{Structure of the equatorial TCOs in the immediately vicinity of a \textit{radially} and \textit{vertically} unstable LR.}
				\label{Fig:StructureFullyUnstable}
			\end{figure}
		
		\item \textit{Vertically} stable.
		
			Here, the allowed TCOs will be \textit{radially} unstable but \textit{vertically} stable.
			
			\begin{figure}[h!]
				\centering
				\includegraphics[scale=0.205]{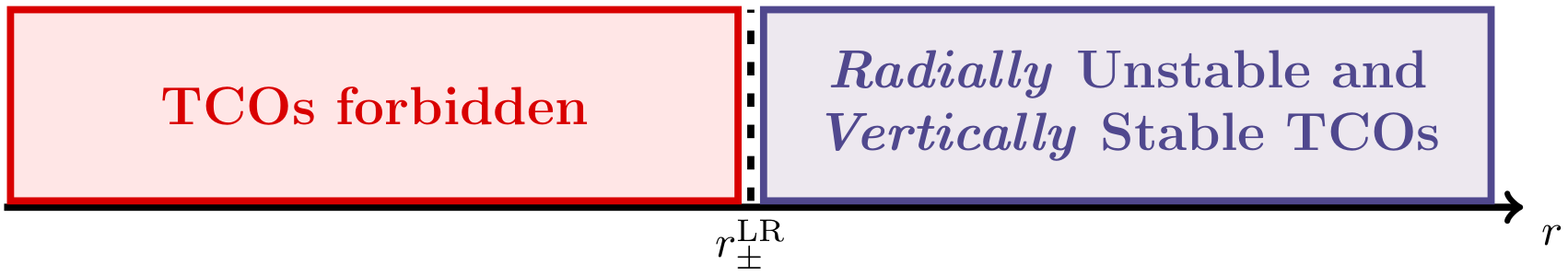}
				\caption{Structure of the equatorial TCOs in the immediately vicinity of a \textit{radially} unstable and \textit{vertically} stable LR.}
				\label{Fig:StructureRadiallyUnstableAngularlyStable}
			\end{figure}
	\end{itemize}
	
	\item Spacetime with a \textit{radially} stable LR.
	
	For such spacetime, the region immediately outwards of the LR can not accommodate TCOs, whereas the inward region can. Similar as the previous case, the full stability of the allowed region is determined by the full stability of the LR. Let us assume that the LR is,
	
	\begin{itemize}
		\item \textit{Vertically} unstable.
		
			For this case, the allowed TCOs with be \textit{radially} stable but \textit{vertically} unstable.
			\begin{figure}[h!]
				\centering
				\includegraphics[scale=0.205]{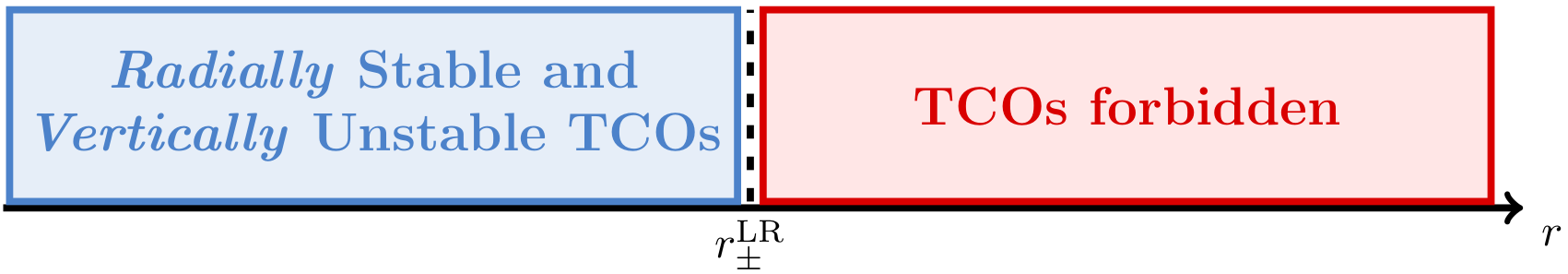}
				\caption{Structure of the equatorial TCOs in the immediately vicinity of a \textit{radially} stable and \textit{vertically} unstable LR.}
				\label{Fig:StructureRadiallyStableAngularlyUnstable}
			\end{figure}
		
		\item \textit{Vertically} stable.
			
			In this particular case, the allowed TCOs will be both \textit{radially} and \textit{vertically} stable.
		
			\begin{figure}[h!]
				\centering
				\includegraphics[scale=0.205]{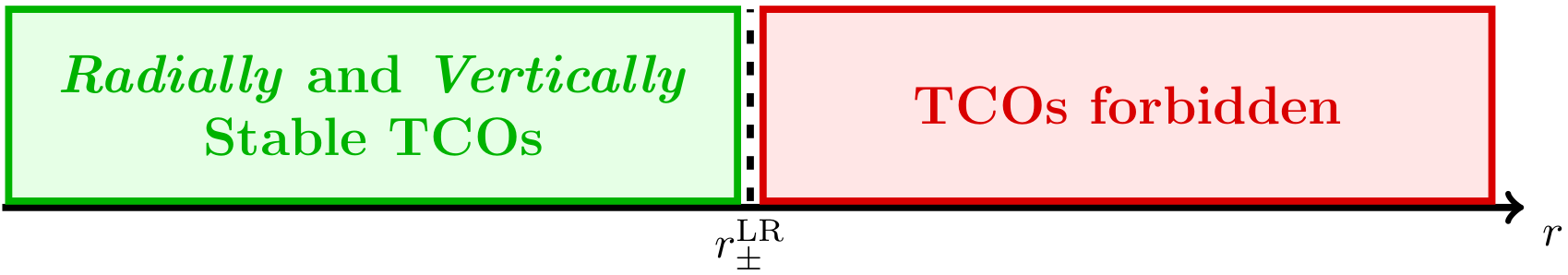}
				\caption{Structure of the equatorial TCOs in the immediately vicinity of a \textit{radially} and \textit{vertically} stable LR.}
				\label{Fig:StructureFullyStable}
			\end{figure}
	\end{itemize}
\end{itemize}




\section{Epicyclic frequencies and energy conditions}

We have showed, so far, that the structures of TCOs around a LR for a generic stationary, axisymmetric and asymptotically flat object with a $\mathbb{Z}_2$ symmetry can be characterised in 4 different ways. This, however, does not assume any type of constrain regarding the matter composing the object. If one wants to consider a object that is compose of real physical matter, then one assumes that the matter obeys the energy conditions.

The energy conditions that we will take special attention to are the NEC and SEC. 
Regarding the former, a proof given in \cite{Cunha:2017qtt} showed that an ultracompact horizonless object, that obeys the NEC, can never have a fully unstable LR. Likewise, following \cite{Cunha:2020azh}, a BH can not have fully unstable LRs if its matter obeys the NEC. Here, we shall arrive to the same result using a different approach than the one used in \cite{Cunha:2017qtt,Cunha:2020azh}. 
Regarding the latter, we shall show a similar conclusion as the one stated previously but related with TCOs. In particular, we shall present a proof that the existence of a region with fully unstable TCOs implies the violation of the SEC.
We will also look into the more simple case of a Ricci-flat object, where we will show that such object can never have neither fully unstable nor stable LRs, and, consequently, can never have neither fully stable or unstable TCOs close to the LRs. However, fully stable TCOs can still exist in regions far away from the LRs, such as, \textit{e.g.}, in the asymptotically flat region.

All previous statement can be demonstrated by relating the sum of the squared of the epicyclic frequencies and the Ricci tensor. 

Following a similar work developed in~\cite{Vieira:2017zau}, where the authors constructed a relation between the epicyclic frequencies and the Ricci tensor, for a generic static, axisymmetric and asymptotically flat spacetime, we need to find a linear combination of the Ricci tensor components such that we can write it in terms of the epicyclic frequencies.
From the connection with the Newtonian limit \cite{Abramowicz:2004tm,binney2011galactic}, we expect that $R_{tt}$ must be connected to the epicyclic frequencies, since, in this limit\cite{Vieira:2017zau,Abramowicz:2004tm,binney2011galactic}, $(\nu_{-1}^r)^2 + (\nu_{-1}^\theta)^2 = 2\Omega_\pm + R_{tt}$, where $R_{tt} = 4\pi G \rho$ -- \textit{cf.} Eq. \eqref{Eq:SumEpicyclicFreqNewtonian}. Furthermore, since the epicyclic frequencies do not depend on the derivatives of $g_{rr}$ and $g_{\theta\theta}$, the linear combination must account that fact. 

The linear combination of Ricci tensor components that satisfy the requirements presented above must be of the form of $R_{tt} + 2 f R_{t\varphi} + f^2 R_{\varphi\varphi}$, where $f$ is either the inverse impact parameter of null particles, $\sigma_\pm$, or the angular velocity of timelike particles, $\Omega_\pm$.
This way, the correct terms to write the epicyclic frequencies appear, and all the terms with derivatives of $g_{rr}$ and $g_{\theta\theta}$ disappear.

Let us focus in the particular case of TCOs. In this case, $f=\Omega_\pm$, and we start by expanding the linear combination, 
\begin{eqnarray}\label{Eq:RicciCombination}
	R_{tt} + 2\Omega_\pm R_{t\varphi} + \Omega_\pm^2 R_{\varphi\varphi} &=& \left(\partial_\mu \Gamma^\mu_{tt} + 2 \Omega_\pm \partial_\mu \Gamma^\mu_{t\varphi} + \Omega_\pm^2 \Gamma^\mu_{\varphi\varphi} \right) + \nonumber \\
	&& \Gamma^\mu_{\mu\nu} \left( \Gamma^\nu_{tt} + 2 \Omega_\pm \Gamma^\nu_{t\varphi} + \Omega_\pm^2 \Gamma^\nu_{\varphi\varphi} \right) - \nonumber \\
	&& \left( \Gamma^\mu_{\nu t} \Gamma^\nu_{\mu t} + 2 \Omega_\pm \Gamma^\mu_{\nu t} \Gamma^\nu_{\mu \varphi} + \Omega_\pm^2 \Gamma^\mu_{\nu \varphi} \Gamma^\nu_{\mu \varphi} \right) ~,
\end{eqnarray}
where $\Gamma^\alpha_{\mu\nu} = \frac{1}{2} g^{\alpha \rho}\left(\partial_\mu g_{\rho \nu} + \partial_\nu g_{\mu \rho} - \partial_\rho g_{\mu\nu} \right)$ are the Christoffel symbols. 
Since we are computing this quantities in a circular orbit at the equatorial plane, we can use the relations discussed in previous sections. This way, the first term (first line) in Eq.  \eqref{Eq:RicciCombination} reduces to,
\begin{equation}
	\partial_\mu \Gamma^\mu_{tt} + 2 \Omega_\pm \partial_\mu \Gamma^\mu_{t\varphi} + \Omega_\pm^2 \Gamma^\mu_{\varphi\varphi} = (\nu_{-1}^r)^2 + (\nu_{-1}^\theta)^2 - \frac{1}{g_{rr}} \frac{C \beta_\pm}{B} ~.
\end{equation}
Likewise, the second term (second line) vanishes, whereas the third and final term (third line) reduces to,
\begin{equation}
	\Gamma^\mu_{\nu t} \Gamma^\nu_{\mu t} + 2 \Omega_\pm \Gamma^\mu_{\nu t} \Gamma^\nu_{\mu \varphi} + \Omega_\pm^2 \Gamma^\mu_{\nu \varphi} \Gamma^\nu_{\mu \varphi} = -\frac{1}{2 g_{rr}} \frac{C \beta_\pm}{B} ~.
\end{equation}
Therefore, we can write the sum of the squared epicyclic frequencies through the linear combination of Ricci tensor components in the following way,
\begin{equation}\label{Eq:SumEpicyclicFrequenciesTCOs}
	(\nu_{-1}^r)^2 + (\nu_{-1}^\theta)^2 = R_{tt} + 2\Omega_\pm R_{t\varphi} + \Omega_\pm^2 R_{\varphi\varphi} + \frac{1}{2g_{rr}} \frac{C \beta_\pm}{B}~.
\end{equation}
We have successfully generalised the result presented in \cite{Vieira:2017zau} for the more general case of a stationary spacetime. In the limit of static spacetimes our result converges to the one obtained in \cite{Vieira:2017zau}. 



The same procedure can be done for the case of LRs. In this case, $f = \sigma_\pm$ and, after expanding and simplifying the linear combination, one can obtain a similar expression as seen in Eq. \eqref{Eq:SumEpicyclicFrequenciesTCOs},
\begin{equation}\label{Eq:SumEpicyclicFrequenciesLRs}
	(\nu_0^r)^2 + (\nu_0^\theta)^2 = R_{tt} + 2\sigma_\pm R_{t\varphi} + \sigma_\pm^2 R_{\varphi\varphi}~.
\end{equation}
Thus, for null particles, the sum of the squared epicyclic frequencies can be entirely written in terms of the Ricci tensor components and the inverse impact parameter.
Note that the same result can be obtained by simply using some results from the previous section, in particular, $\beta_\pm = 0$, $\sigma_\pm = \Omega_\pm$, $\nu_{-1}^r = \nu_0^r$ and $\nu_{-1}^\theta = \nu_0^\theta$, and using them in Eq. \eqref{Eq:SumEpicyclicFrequenciesTCOs}.

We can now investigate what happen to Eqs. \eqref{Eq:SumEpicyclicFrequenciesTCOs} and \eqref{Eq:SumEpicyclicFrequenciesLRs} when we impose some conditions to the Ricci tensor.


\subsection{Ricci-flat object}

For a Ricci-flat object, all components of the Ricci tensor vanish, $R_{\mu\nu} = 0$. For such objects, the sum of the squared epicyclic frequencies of null and timelike particles simplifies to,
\begin{eqnarray}
	\text{LR:}&~ (\nu_0^r)^2 + (\nu_0^\theta)^2 = 0~, \label{Eq:RicciFlatLRs} \\
	\text{TCOs:}&~ (\nu_{-1}^r)^2 + (\nu_{-1}^\theta)^2 = \dfrac{1}{2g_{rr}} \dfrac{C \beta_\pm}{B}~. \label{Eq:RicciFlatTCOs}
\end{eqnarray}

From the first equation, one can conclude that a LR can never be neither fully stable, $(\nu_0^r)^2 > 0 ~ \land ~ (\nu_0^\theta)^2 > 0$, or unstable, $(\nu_0^r)^2 < 0 ~ \land ~ (\nu_0^\theta)^2 < 0$, since its sum must vanish. This implies that the LR must have opposite radial and vertical stabilities. 
This is consistent with what is already known for Ricci-flat solutions that are stationary, axisymmetric, asymptotically flat and in 1+3 dimensions, such as the Kerr solution~\cite{abramowicz1998theory,abramowicz2013foundations,Torok:2005ct}. Both the prograde and retrograde LRs present in the Kerr spacetime are always radially unstable and vertically stable. 

For the case of TCOs, one can prove that the right-hand side of Eq. \eqref{Eq:RicciFlatTCOs} is always positive. The proof rely on the fact that each individual quantity is positive. The $g_{rr}$ component of the metric and the $B$ function are always positive when we are outside of a possible horizon -- \textit{cf.} Sec. \ref{Sec:Definitions}. The function $C$ is always positive on the regions where TCOs are allowed, otherwise the angular velocity of the timelike particle is complex, and we no longer have TCOs -- see \cite{Delgado:2021jxd} for a more detailed analysis. The $\beta_\pm$ function is also always positive on the regions where TCOs are allowed, as we discussed in the previous section. Therefore, we have proved that the right-hand size of Eq. \eqref{Eq:RicciFlatTCOs} is always positive on the regions where we can find TCOs.

A direct consequence of this proof is that TCOs around a Ricci-flat solution can never be fully unstable. However, their stability can be any of the three remaining possibilities combinations. For the well-known case of Kerr BHs, it is known that sufficiently far away from the BH, TCOs are fully stable, whereas, if we approach the BH, we start to find radially unstable but vertically stable TCOs~\cite{abramowicz1998theory,abramowicz2013foundations,Torok:2005ct}. Thus, our result is consistent with the well-known results from the Kerr case.

\subsection{NEC and SEC obeying object}

If an object possesses matter such that its energy-momentum tensor (and, by Einstein's equations, the Ricci tensor) is non-vanishing, a way to inform ourselves about its exotic properties is by evaluating possible violation of the energy conditions. 
One of them is the SEC. In short, this condition is defined as $R_{\mu\nu} t^\mu t^\nu \geq 0$, for any timelike vector field, $t^\mu$, and encapsulated the condition that matter must gravitate towards matter.   
To study this energy condition, let us consider a timelike vector tangent to a TCO,
\begin{equation}
	t^\mu = a \left( \eta_1^\mu + \Omega_\pm \eta_2^\mu \right)~,
\end{equation}
where $a = a(r_\text{cir},\pi/2) > 0$ is a constant. By computing the SEC, we arrive to,
\begin{equation}
	R_{\mu\nu} t^\mu t^\nu \geq 0 ~~\Leftrightarrow~~ a^2 \left( R_{tt} + 2 \Omega_\pm R_{t\varphi} + \Omega_\pm^2 R_{\varphi\varphi} \right) \geq 0~.
\end{equation}
Therefore, we can written the sum of the squared epicyclic frequencies of a TCO as,
\begin{equation}\label{Eq:SECObeyingTCOs}
	(\nu_{-1}^r)^2 + (\nu_{-1}^\theta)^2 = \frac{1}{a^2} R_{\mu\nu} t^\mu t^\nu + \dfrac{1}{2g_{rr}} \dfrac{C \beta_\pm}{B}~. 
\end{equation}
Thus, for a object whose matter obeys the SEC, the first term of the right-hand size of Eq. \eqref{Eq:SECObeyingTCOs} is always positive. Furthermore, using the same argument used in the previous subsection where we prove that the right-hand size of Eq. \eqref{Eq:RicciFlatTCOs} is always positive, we can conclude that the sum of the squared epicyclic frequencies of TCOs around an object obeying the SEC is always positive. 
This implies that no fully unstable TCOs exist on the spacetime generated by such object.
Or, in different words, if a generic stationary, axisymmetric, asymptotically flat spacetime has a region with fully unstable TCOs, then we can guarantee that the SEC is violated.

A second energy condition that we can also study is the NEC. This condition is defined as $T_{\mu\nu} k^\mu k^\nu \geq 0$, for any null vector field $k^\mu$, and it is intertwined with the SEC, in the sense that one can imply the other. 
In particular, if the SEC is obeyed then the NEC is also obeyed, or, if the NEC is violated, the SEC is also violated. 
We will now show that the interlacement between both energy conditions is consistent with all results discussed so far. 

Let us consider the following null vector field,
\begin{equation}
	k^\mu = b \left( \eta_1^\mu + \sigma_\pm \eta_2^\mu \right)~,
\end{equation}
where $b = b(r_\text{cir},\pi/2) > 0$. With this null vector, the NEC can be written as\footnote{We are assuming that, within a given theory of gravity, we can defined a effective energy-momentum tensor such that we can use the Einstein's equations, $R_{\mu\nu} - \frac{1}{2}g_{\mu\nu} R = T_{\mu\nu}^\text{eff}$.},
\begin{equation}
	T_{\mu\nu} k^\mu k^\nu \geq 0 ~~\Leftrightarrow~~ b^2 (R_{tt} + 2\sigma_\pm R_{t\varphi} + \sigma_\pm^2 R_{\varphi\varphi}) \geq 0 ~.
\end{equation}
Therefore, the sum of the squared epicyclic frequencies of the LRs take a very simple form,
\begin{equation}\label{Eq:SECObeyingLRs}
	(\nu_0^r)^2 + (\nu_0^\theta)^2 = \frac{1}{b^2} T_{\mu\nu} k^\mu k^\nu~. 
\end{equation}
From this result we can conclude that fully unstable LRs never exist for objects whose matter obeys the NEC. 
Or, in different words, if a generic stationary, axisymmetric, asymptotically flat spacetime has fully unstable LRs, then we can guarantee that the NEC is violated.
This is precisely the same result that was first obtained in \cite{Cunha:2017qtt} for an ultracompact horizonless object and \cite{Cunha:2020azh} for a BH. 
 
Finally, let us comment on the interlacement of the energy conditions and all results discussed so far. Consider now that we have a hypothetical spacetime such that it possesses a fully unstable LR, hence it violated the NEC. From the result obtained in the previous section, we know that in the immediately vicinity of the LR we have a region which harbours fully unstable TCOs -- \textit{cf.} Fig. \ref{Fig:StructureFullyUnstable}. Thus, by Eq. \eqref{Eq:SECObeyingTCOs} and the following argument, the SEC is also violated. 
This is consistent with the implication tree of the NEC and SEC. Namely, that the violation of the NEC also implies the violation of the SEC.

\section{Conclusions and final remarks}

In this work, we have analysed the radial and vertical epicyclic frequencies of null and timelike particles following circular orbits in a spacetime generated by a generic stationary, axisymmetric, asymptotically flat ultracompact object with a $\mathbb{Z}_2$ symmetry. We have shown that the epicyclic frequencies of LRs coincide with the epicyclic frequencies of TCOs when the latter approach the former. This implies, by continuity, that the full stability of a LR determines the full stability of TCOs in the immediately vicinity of the LR.
The results present in the first part of this paper generalise the ones obtain in \cite{Delgado:2021jxd}, where the authors only analysed the radial stability of the LRs and TCOs. 
Together with the results of \cite{Delgado:2021jxd}, we determined the 4 possible structures that one can found for a generic spacetime with a LR -- \textit{cf.} Figs. \ref{Fig:StructureFullyUnstable} -- \ref{Fig:StructureFullyStable}. 

The results of the first part of this paper can be summarised as follows: the \textit{radial} stability of a LR determines the localisation and \textit{radial} stability of TCOs in its vicinity, whereas, the \textit{vertical} stability of a LR only determines the \textit{vertical} stability of TCOs in its vicinity.

In the second part of this work, we established relations between the sum of the squared epicyclic frequencies with linear combinations of the Ricci tensor components, that generalise the results obtained in \cite{Vieira:2017zau}. Such relations opened the possibility of introduce some energy conditions regarding the matter that the generic ultracompact object may be compose of. In particular, we showed that if the object obeys the SEC, which implies that the NEC is also obeyed, LRs and TCOs can never be fully unstable. Reciprocally, if a generic ultracompact object possesses fully unstable LRs, it violated the NEC, and consequently the SEC. This is consistent with the works presented in \cite{Cunha:2017qtt,Cunha:2020azh}. Furthermore, if the same object only possesses a region with fully unstable TCOs, then the object only violates the SEC. 

We also study the more simpler case of a Ricci-flat ultracompact object. In this case, we showed that the LRs can never be neither fully stable or fully unstable, since the sum of the squared epicyclic frequencies of each LR always vanishes -- \textit{cf.} Eq. \eqref{Eq:RicciFlatLRs}. Regarding the TCOs, the sum of its epicyclic frequencies is always positive, hence, a Ricci-flat object never has regions of fully unstable TCOs.

As future work, it would be interesting to generalise further the results presented here to more generic ultracompact objects that do not have a $\mathbb{Z}_2$ symmetry. For such objects, circular orbits will not always be found at the equatorial plane, but will be found at an angular coordinate $\theta_\text{cir}$ that will be a function of the radial coordinate, $\theta_\text{cir} = f(r_\text{cir})$. Due to the missing $\mathbb{Z}_2$ symmetry, the relation between the full stability of a LR and the full stability of TCOs in its vicinity will, most likely, be more convoluted and less obvious. Nevertheless, one could obtain all the possible structures of TCOs around LRs and also study which are connected to violations of the energy conditions.


\section{Acknowledgements}

We would like to thank C. Herdeiro and E. Radu for comments on a draft of this paper. 
This work is supported by the Center for Research and Development in Mathematics and Applications (CIDMA) and Center for Astrophysics and Gravitation (CENTRA) through the Portuguese Foundation for Science and Technology (FCT - Fundação para a Ciência e a Tecnologia), references UIDB/04106/2020, UIDP/04106/2020 and UIDB/00099/2020. 
We acknowledge support from the projects PTDC/FIS-OUT/28407/2017, CERN/FISPAR /0027/2019, 
PTDC/FIS-AST/3041/2020. This work has further been supported by the European Union’s Horizon 2020 research and innovation (RISE) programme H2020-MSCA-RISE-2017 Grant No. FunFiCO-777740.


\bibliographystyle{ieeetr}
\bibliography{bibliography}

\end{document}